\documentclass[prl,aps,twocolumn,amsfonts,showpacs]{revtex4}
\usepackage{epsfig} 
\usepackage{psfrag} 
\usepackage{amsmath}
\usepackage{amssymb} 
\usepackage{color} 
\usepackage{graphicx}
\usepackage{subfigure}

\begin{document}

\title{Feshbach Resonance in Optical Lattices and the Quantum Ising Model} 
\author{M. J. Bhaseen} 
\affiliation{University of
Cambridge, Cavendish Laboratory, Cambridge, CB3 0HE, UK.}
\author{A. O. Silver} 
\affiliation{University of
Cambridge, Cavendish Laboratory, Cambridge, CB3 0HE, UK.}
\author{M. Hohenadler}
\altaffiliation{Present Address: OSRAM Opto 
Semiconductors GmbH, 93055 Regensburg, GER.}
 \author{B. D. Simons} 
\affiliation{University of
Cambridge, Cavendish Laboratory, Cambridge, CB3 0HE, UK.}
\altaffiliation{Present Address: OSRAM Opto 
Semiconductors GmbH, 93055 Regensburg, GER.}

\begin{abstract} 
  Motivated by experiments on heteronuclear Feshbach resonances in
  Bose mixtures, we investigate $s$-wave pairing of two species of
  bosons in an optical lattice. The zero temperature phase diagram
  supports a rich array of superfluid and Mott phases and a network of
  quantum critical points.  This topology reveals an underlying
  structure that is succinctly captured by a two-component Landau
  theory. Within the second Mott lobe we establish a quantum phase
  transition described by the paradigmatic longitudinal and transverse
  field Ising model.  This is confirmed by exact diagonalization of
  the 1D bosonic Hamiltonian.  We also find this transition in the
  homonuclear case.
\end{abstract}


\maketitle

{\em Introduction.}--- With the advent of ultra-cold atomic gases, the
fermionic BEC--BCS crossover has been the focus of intense scrutiny
\cite{Donley:Atmol,Regal:Creation,Regal:BCSBEC}. Tremendous
experimental control has been achieved through the use of Feshbach
resonances, which allow one to manipulate atomic interactions by a
magnetic field. By sweeping the strength of attraction, one may
interpolate between a BEC of tightly bound molecules, and a BCS state
of loosely associated pairs.  More recently, the analogous problem for
a single species of boson has been the subject of theoretical
investigation
\cite{Rad:Atmol,Romans:QPT,Koetsier:Bosecross,Radzi:Resonant,Rousseau:Fesh,Rousseau:Mixtures}.
An important distinction from the fermionic case is that the carriers
themselves may condense.  This leads to the possibility of novel
phases and phase transitions, with no fermionic counterpart.

In parallel, there has been a significant experimental drive to study
Feshbach resonances in binary mixtures of different atomic species.
An important catalyst is the quest for heteronuclear molecules as a
route to dipolar interactions and exotic condensates
\cite{Santos:Dipolar}.  Recently, heteronuclear molecules have been
created in $^{85}{\rm Rb}$--$^{87}{\rm Rb}$ bosonic mixtures through
both s-wave \cite{Papp:Hetero} and p-wave resonances
\cite{Papp:Tunable,Ticknor:Multiplet}. Similarly, bosonic $^{41}{\rm
  K}$--$^{87}{\rm Rb}$ mixtures have been studied in harmonic
potentials \cite{Modugno:Two} and in optical lattices
\cite{Catani:Bosebose}, and interspecies resonances have also been
achieved \cite{Weber:Assoc,Thalhammer:Double,Thalhammer:Coll}.  The
enhanced longevity of molecules in optical lattices
\cite{Thalhammer:Long}, and the sympathetic cooling of one species by
the other, make these attractive for experiment. Multiple species also
provide additional ``isospin'' degrees of freedom, and offer exciting
possibilities for interesting phases and quantum magnetism
\cite{Altman:Twocomp}.

Motivated by these significant developments we consider the s-wave
heteronuclear Feshbach problem for two-component bosons in an optical
lattice. Our primary goal is to establish and explore the rich phase
diagram, which supports distinct atomic and molecular superfluids in
proximity to Mott phases. We also shed light on the nature of the
quantum phase transitions, and a key finding is a transition described
by the paradigmatic quantum Ising model. This ubiquitous model plays a
central role in a variety of quantum many body contexts, and a
controllable realization in cold atomic gases may open new directions
on dynamics and frustrated lattices.

{\em The Model.}--- Let us consider a two-component Bose gas with a
``spin'' index $\downarrow,\uparrow$.  We assume that these components
may form molecules, $m$. The Hamiltonian
\begin{equation}
\begin{aligned}
H & =\sum_{i\alpha}\epsilon_\alpha n_{i\alpha}- \sum_{\langle
ij\rangle} \sum_\alpha t_\alpha\left(a_{i\alpha}^\dagger
a_{j\alpha}+{\rm h.c.} \right)\\ & \hspace{1cm}
+\sum_{i\alpha\alpha^\prime}
\frac{U_{\alpha\alpha^\prime}}{2}:n_{i\alpha}n_{i\alpha^\prime}: +
\,H_{F},
\label{atmolham}
\end{aligned}
\end{equation}
describes bosons, $a_{i\alpha}$, hopping on a lattice with sites $i$,
where $\alpha=\downarrow,\uparrow,m$ labels the carrier. Here,
$\epsilon_\alpha$ are onsite potentials, $t_\alpha$ are hopping
parameters, $\langle ij\rangle$ denotes summation over nearest
neighbor bonds, and $U_{\alpha\alpha^\prime}$ are interactions. We
assume that molecule formation is described by the s-wave interspecies
Feshbach resonance term
\begin{equation}
H_{F}= g\sum_i(a^\dagger_{im} a_{i\uparrow}a_{i\downarrow}+{\rm h.c.}).
\label{HF}
\end{equation}
For recent work on the $p$-wave problem see Ref.~\cite{Radzi:spinor}.
Normal ordering yields $:n_{i\alpha}n_{i\alpha}: =
n_{i\alpha}(n_{i\alpha}-1)$ for like species, and
$:n_{i\alpha}n_{i\alpha^\prime}:=n_{i\alpha}n_{i\alpha^\prime}$ for
distinct species. To aid numerical simulations we consider hardcore
atoms and molecules and set $U_{m\uparrow}=U_{m\downarrow}\equiv U$
and $U_{\uparrow\downarrow}\equiv V$. We work in the grand canonical
ensemble with $H_\mu=H-\mu_{\rm T} N_{\rm T}-\mu_{\rm D}N_{\rm D}$,
where $N_{\rm T}\equiv \sum_i
(n_{i{\uparrow}}+n_{i\downarrow}+2n_{im})$ and $N_{\rm D}\equiv
\sum_i(n_{i\uparrow}-n_{i\downarrow})$ commute with $H$.  These are
the total atom number (including a factor of two for molecules) and
the up-down population imbalance respectively.

\begin{figure} 
\includegraphics[width=6cm]{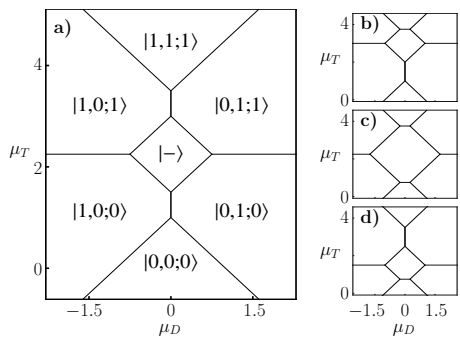}
\caption{Zero hopping phase diagram showing eigenstates in the
  occupation basis $|n_\downarrow,n_\uparrow;n_m\rangle$ with
  $\epsilon_\downarrow=\epsilon_\uparrow=U=1$, $V=1.5$ and (a) $g=1$,
  $\epsilon_m=3.5$, (b) $g=1$, $\epsilon_m=5$, (c) $g=2$,
  $\epsilon_m=3.5$, (d) $g=1$, $\epsilon_m=2$.}
\label{Fig:Zerohop}
\end{figure}

{\em Phase Diagram.}--- To establish the phase diagram it is
convenient to first examine the zero hopping limit. This helps
constrain the overall topology and provides some orientation for the
general problem. With three species of hardcore bosons we need to
consider eight possible states in the occupation basis,
$|n_\downarrow,n_\uparrow;n_m\rangle$.  The Feshbach coupling, $g$,
only mixes $|1,1;0\rangle$ and $|0,0;1\rangle$, and the resulting
eigenstates $|\pm\rangle$ have energies
$E_\pm=\tilde\epsilon_m-h/2\pm\sqrt{(h/2)^2+g^2}$, where $h\equiv
\epsilon_m-\epsilon_\downarrow-\epsilon_\uparrow-V$ and the chemical
potentials are absorbed into
$\tilde\epsilon_\uparrow\equiv\epsilon_\uparrow-\mu_{\rm T}-\mu_{\rm
  D}$, $\tilde\epsilon_\downarrow\equiv\epsilon_\downarrow-\mu_{\rm
  T}+\mu_{\rm D}$, $\tilde\epsilon_m\equiv\epsilon_m-2\mu_{\rm T}$.
The remaining six states have energies
$E(n_\downarrow,n_\uparrow;n_m)=\sum_\alpha\tilde\epsilon_\alpha
n_\alpha+Vn_\uparrow
n_\downarrow+Un_m(n_\uparrow+n_\downarrow)$. Minimizing over all
states we obtain the zero hopping diagram shown in
Fig.~\ref{Fig:Zerohop}. In analogy with the single band Bose--Hubbard
model \cite{Fisher:Bosonloc} the total density $\rho_{\rm T}\equiv
n_\downarrow+n_\uparrow+2n_m$ is pinned to integer values and
increases with $\mu_{\rm T}$. Increasing (decreasing) $\mu_{\rm D}$
favors up (down) atoms as is evident from the definition of $N_{\rm
  D}$.
In contrast to the single band case \cite{Fisher:Bosonloc}, the
topology of the diagram changes as a function of the parameters. In
particular, the shape and extent of the central Mott lobe $|-\rangle$
depends on the strength of the Feshbach coupling, $g$, and may
terminate directly with the vacuum state and/or the completely filled
state as shown in Fig. \ref{Fig:Zerohop} (b)-(d).

\begin{figure}
\begin{center}
\includegraphics[clip,width=7.8cm]{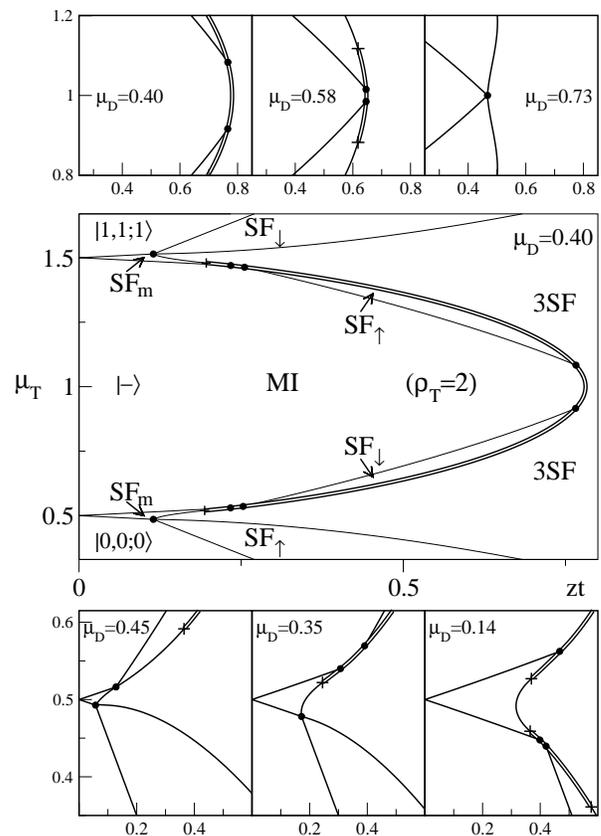}
\end{center}
\caption{Center: Mean field phase diagram for $\mu_{\rm D}=0.4$,
$\epsilon_\uparrow=\epsilon_\downarrow=1$, $\epsilon_m=2$, $g=1$,
$U=V=0$, $t_\uparrow=t_\downarrow=t$, and
$t_m=t/2$. We indicate the one-component up, down and molecular
superfluids, by ${\rm SF}_\uparrow$, ${\rm SF}_\downarrow$ and ${\rm
SF}_m$, while MI denotes the Mott insulating phase. 
Phase 3SF has all three components superfluid. We denote first
order (continuous) transitions by double (single) lines. Junctions
between phases are indicated by a dot, and the termination of first
order lines by a cross.  Top: Magnified portion of the central lobe
tip showing the retreat of the first order transitions and the
emergence of a tetracritical point.  Bottom: Magnified portion of the
lower left region showing the underlying tetracritical points as
$\mu_D$ is varied.
\label{Fig:MFT}}
\end{figure}

We now consider the effect of the hopping terms. To decouple these we
make the mean field ansatz $\phi_\alpha\equiv \langle
a_{i\alpha}\rangle$ for each component, and replace
$a_{i\alpha}\rightarrow \phi_\alpha+(a_{i\alpha}-\phi_\alpha)$. This
yields the effective single site Hamiltonian
\begin{equation}
H=H_0-\sum_\alpha zt_\alpha\left(a_\alpha^\dagger\phi_\alpha 
+\phi_\alpha^\ast a_\alpha-
|\phi_\alpha|^2\right),
\label{MFTham}
\end{equation}
where $H_0$ is the single site zero hopping contribution to
(\ref{atmolham}), and $z$ is the coordination. We minimize
(\ref{MFTham}) to obtain the phase diagram in Fig. \ref{Fig:MFT},
where the symmetry under $\mu_{\rm T}\rightarrow 2-\mu_{\rm T}$
reflects invariance of the Hamiltonian upon particle-hole and
interchange operations, $a_\alpha\leftrightarrow a_\alpha^\dagger$,
$\mu_{\rm T}\rightarrow \epsilon_m+U-\mu_{\rm T}$,
$a_\downarrow\leftrightarrow a_\uparrow$, when
$t_\downarrow=t_\uparrow$ and $h=0$.

The phase diagram has a rich structure and exhibits single component
atomic and molecular condensates, and a region with all three
superfluid.  A notable absence is a phase where just two components
are superfluid. This is a consequence of the Feshbach term (\ref{HF});
condensation of any two variables acts like an effective field on the
remaining species and induces three component superfluidity.  In
contrast, condensation of a single variable no longer acts like a
field and single component superfluids are supported.  For example,
with $\mu_{\rm D}>0$ we have
$\tilde\epsilon_\downarrow>\tilde\epsilon_\uparrow$ and on leaving the
vacuum we enter either the up or molecular superfluid. The former is
favored at large hopping due to the chosen hopping asymmetry; see
Fig.~\ref{Fig:MFT}.

The appearance of single component atomic superfluids, with $\langle
a_\downarrow\rangle\neq 0$ or $\langle a_\uparrow\rangle \neq 0$,
distinguishes this from the homonuclear case
\cite{Rad:Atmol,Romans:QPT}.  In the latter, single component atomic
superfluids are absent due to the reduced form of the 
Feshbach term $\sim g(a_m^\dagger a a+{\rm h.c.})$. This difference
also shows up in the symmetry classification of the phase transitions.
In the heteronuclear problem defined by equation (\ref{HF}), molecular
condensation leaves a ${\rm U}(1)$ symmetry intact ($a_{i\downarrow
}\rightarrow e^{i\alpha} a_{i\downarrow}$, $a_{i\uparrow}\rightarrow
e^{-i\alpha}a_{i\uparrow}$) in contrast to the ${\mathbb Z}_2$
symmetry ($a\rightarrow -a$) of Refs. \cite{Rad:Atmol,Romans:QPT}.
The transition from the molecular to three component superfluid is
thus expected to be in the XY universality class along its continuous
segment rather than Ising.

{\em Landau Theory.}--- The phase diagram in Fig.~\ref{Fig:MFT}
displays an elaborate network of quantum critical points and phase
transitions. This topology reveals an underlying structure that is
succinctly captured by Landau theory.  In the absence of competition
from other phases the locus of continuous transitions from the Mott
states to the one-component superfluids may be determined
analytically. This may be done using second order perturbation theory
to locate the $|\phi_\alpha|^2$ terms, or through diagonalization of
(\ref{MFTham}) when two $\phi_\alpha$ are set to zero; e.g.\ the
transition from $|0,0;0\rangle$ to the up-superfluid with $\langle
a_{i\uparrow}\rangle\equiv \phi_\uparrow\neq 0$ occurs along a segment
of $zt_\uparrow=\tilde\epsilon_\uparrow$.  More generally,
\begin{equation}
\begin{aligned}
  E = E_0 & +\frac{1}{2}\sum_{\alpha}m_\alpha|\phi_\alpha|^2+
  \frac{\gamma}{2}\left(\phi_m^\ast\phi_\downarrow\phi_\uparrow+{\rm
      h.c.}\right)\\ & +
  \frac{1}{4}\sum_{\alpha\beta}\lambda_{\alpha\beta}|\phi_\alpha|^2|\phi_{\beta}|^2+{\mathcal
    O}(\phi^6),
\end{aligned}
\label{LT}
\end{equation}
where $\lambda_{\alpha\alpha}>0$, and the detailed form of the
coefficients (but not the structure) depend on the unperturbed Mott
state with energy $E_0$.  This is a Bose--Hubbard ${\rm U}(1)$ Landau
theory for each component \cite{Fisher:Bosonloc}, supplemented by
couplings allowed by the ${\rm U}(1)\times {\rm U}(1)$ symmetry of
(\ref{atmolham}). A similar model (without permissible biquadratic
terms) was put forward on phenomenological grounds to describe pairing
in the two-band Bose--Hubbard model {\em without} a Feshbach term
\cite{Kuklov:Commens}. One may gain insight into the Landau theory
(\ref{LT}) by reduction. Near the lower tetracritical point in
Fig.~\ref{Fig:MFT} for example, $\phi_\downarrow$ is massive
($m_\downarrow>0$) and may be eliminated using its saddle point:
\begin{equation}
  E = E_0+\frac{1}{2}\sum_{\alpha} m_\alpha|\phi_\alpha|^2
  +\frac{1}{4}\sum_{\alpha\beta}\Lambda_{\alpha\beta}|\phi_\alpha|^2|\phi_\beta|^2+\dots,
\label{redlan}
\end{equation}
where $\alpha,\beta=\uparrow,m$,
$\Lambda_{\alpha\alpha}=\lambda_{\alpha\alpha}$ and $\Lambda_{\uparrow
  m}=\lambda_{\uparrow m}-\gamma^2/m_\downarrow$. The behavior of this
reduced Landau theory (\ref{redlan}) depends on the sign and magnitude
of $\Lambda_{\uparrow m}$. For
$\Lambda_{\uparrow\uparrow}\Lambda_{mm}>\Lambda_{\uparrow m}^2$ it
yields a tetracritical point \cite{Chaikin:Book}, whilst away from
this, and for $\Lambda_{\uparrow m}<0$, it yields two tricritical
points separated by a first order transition.  This evolution is borne
out in Fig.~\ref{Fig:MFT}, where we track the development of the phase
diagram as a function of $\mu_{\rm D}$. Indeed, the whole phase
diagram may be understood within this reduced framework as the
evolution of three such tetracritical points and their particle--hole
reflections by eliminating $\phi_\downarrow$,
$\phi_\uparrow$, and $\phi_m$ in turn.

\begin{figure}
\begin{center}
\includegraphics[clip,width=8cm]{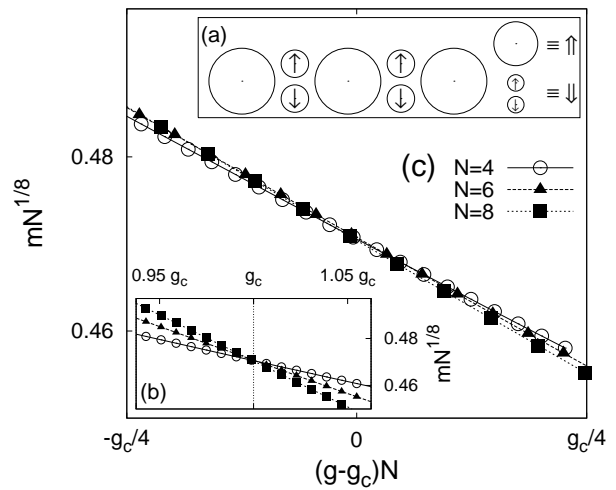}
\caption{(a) A pair of atoms, $a_\downarrow,a_\uparrow$, is
  represented by spin down $|\!\Downarrow\rangle$ and a molecule,
  $a_m$, by spin up $|\!\Uparrow\rangle$.  Hopping favors N\'eel
  order, whilst the Feshbach term acts like a transverse field.  (b)
  $mN^{1/8}$ versus $g$ obtained from exact diagonalization of the 1D
  bosonic model (\ref{atmolham}) for different system sizes, $N$ at
  density $\rho_{\rm T}=2$.  We take
  $\epsilon_\downarrow=\epsilon_\uparrow=1$ and $\epsilon_m=2$ ($h=0$)
  and $U=1$, $V=0$. We set $t=t_\uparrow=t_\downarrow=2t_m=0.01$.  The
  crossing point, $g_c\approx 1.041\times 10^{-4}$, lies slightly
  below the critical value, $g_c=J/4$, of the Ising model
  (\ref{LTFI}).  (c) Scaling collapse as a function of $(g-g_c)N$ with
  the critical exponents $\beta=1/8$ and $\nu=1$ of the 2D classical
  Ising model. }
\label{Fig:Isingexp}
\end{center}
\end{figure}

{\em Mott Phases.}--- Having surveyed the overall phase diagram we
turn our attention to the Mott states. This will reveal Ising
transitions in both the heteronuclear {\em and} homonuclear lattice
problems, unreported in Refs. \cite{Rousseau:Fesh,Rousseau:Mixtures}.
To explore the central lobe where the Feshbach term is operative, we
adopt a magnetic description.  With a pair of atoms or a molecule at
each site, we introduce effective spins $|\!\Downarrow\rangle \equiv
|1,1;0\rangle$, $|\!\Uparrow\rangle \equiv |0,0;1\rangle$; see
Fig.~\ref{Fig:Isingexp}~(a). The operators $S^+=a_m^\dagger a_\uparrow
a_\downarrow$, $S^-=a_\downarrow^\dagger a_\uparrow^\dagger a_m$, and
$S^z=\left(n_m-n_\downarrow n_\uparrow\right)/2$, form a
representation of ${\rm su}(2)$ on this reduced Hilbert space, and
deep within the Mott phase we perform a strong coupling $t/U$
expansion \cite{Duan:Control}:
\begin{equation}
H= J\sum_{\langle ij\rangle}S_i^zS_j^z+\sum_i\left(hS_i^z+\Gamma
S_i^x\right),
\label{LTFI}
\end{equation}
where $\Gamma=2g$ and
$J=2(\frac{t_\downarrow^2+t_\uparrow^2}{U-V}+\frac{t_m^2}{2U})>0$ is
antiferromagnetic exchange.  We have omitted the constant,
$\tilde\epsilon_m-h/2-Jz/8$ per site.  This takes the 
form of a quantum Ising model in a longitudinal and transverse field.
The longitudinal field, $h$, reflects the energetic asymmetry between
a molecule $|\!\Uparrow\rangle$, and a pair of atoms
$|\!\Downarrow\rangle$.  The transverse field, $\Gamma\equiv 2g$,
encodes Feshbach conversion; see Fig.~\ref{Fig:Isingexp}
(a). 
Since XY exchange involves interchanging {\em two} atoms {\em and} a
molecule it enters at $t^3/U^2$ and may be neglected. 

\begin{figure}
\begin{center}
\includegraphics[clip=true,width=8cm]{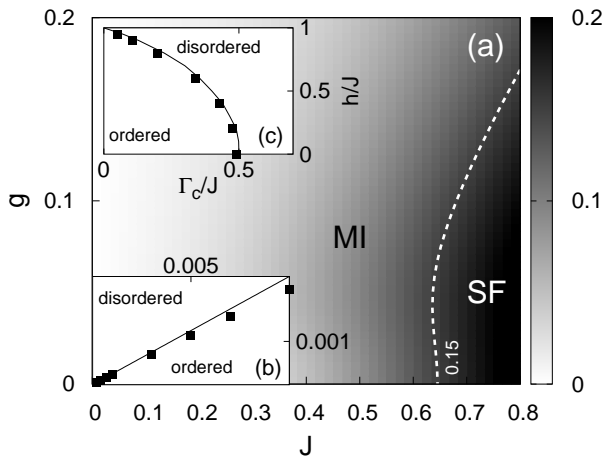}
\caption{(a) Total SF fraction, $f_s$ \cite{Roth:Twocolor}, for $N=8$
  sites and the parameters used in Fig.~\ref{Fig:Isingexp}. The
  contour indicates the approximate location of the MI-SF boundary.
  (b) $g$ versus $J$ showing the Ising transition $g_c(J)$ in the
  second lobe of the 1D bosonic model (\ref{atmolham}) using exact
  diagonalization. In the small $J$ limit, $g_c=J/4$ (solid line) in
  accordance with the transverse field Ising model. (c) $h$-dependence
  of $\Gamma_c=2g_c$ at fixed $J$, for the 1D bosonic (squares) and
  quantum Ising models (solid line). Here $\epsilon_m=2+h$ and all
  other parameters as in Fig.~\ref{Fig:Isingexp}.\label{Fig:Boundary}}
\end{center}
\end{figure}

{\em Numerical Simulations.}--- The model (\ref{LTFI}) is of
considerable importance in a variety of contexts, and underpins much
of our understanding of quantum magnetism and quantum phase
transitions.  To verify this realization in our {\em bosonic} model,
we perform exact diagonalization on the 1D quantum system
(\ref{atmolham}) with periodic boundaries.  At present numerical
techniques are considerably less advanced for multicomponent bosonic
systems, and the large Hilbert space $\propto 2^{3N}$ restricts our
simulations to $N=8$ sites.  We begin with $h=0$ before exploring
finite fields. Owing to the absence of spontaneous symmetry breaking
in finite size systems, the staggered magnetization vanishes. Instead,
it is convenient to focus on $m\equiv \langle |\sum_i
(-1)^iS_i^z|\rangle/N$ \cite{Um:FiniteTFI}, where
$S_i^z=\left[n_{im}-(n_{i\uparrow}+n_{i\downarrow})/2\right]/2$.
Adopting the finite size scaling form, $m=N^{-\beta/\nu} \bar
m\left[(g-g_c)N^{1/\nu}\right]$ \cite{Um:FiniteTFI}, we plot
$mN^{1/8}$ versus $g$ for different system sizes, $N$; see
Fig.~\ref{Fig:Isingexp}~(b). These cross close to the critical
coupling, $g_c=J/4$, of the purely transverse field Ising model, and
the scaling collapse is consistent with the critical exponents
$\beta=1/8$ and $\nu=1$ for the 2D classical model.  Repeating this we
may track the transition within the Mott lobe. The results in
Fig.~\ref{Fig:Boundary}~(b) show clear Ising behaviour at small
hoppings.
In addition, the finite size corrections to the ground state energy
yield the central charge, $c\approx 0.51$, consistent with the exact
result, $c=1/2$.  At larger $J$ we continue to identify a transition
in the Ising universality class, but the locus is modified. Our system
sizes are insufficient to track this all the way to the superfluid
boundary; see Fig.~\ref{Fig:Boundary}.  These issues will be addressed
in detail elsewhere by DMRG.  Having established an Ising transition
at $h=0$, we now consider $h\neq 0$. As shown in
Fig.~\ref{Fig:Boundary}~(c), the location of the critical point
evolves in accordance with exact diagonalization of (\ref{LTFI}) and
DMRG results on the Ising model \cite{Ovchinnikov:Mixed}. The leading
quadratic correction to the ground state energy, $\delta
E_0^{(2)}\approx -0.07h^2N$ at $g=g_c$ \cite{Ovchinnikov:Mixed} is
also compatible \footnote{The {\em ferromagnet} with ${\rm E}_8$
  spectrum has $\delta E_0 \propto h^{8/15}$.}.  This confirms an
Ising transition in our {\em bosonic} model for a range of parameters
without fine tuning. This also occurs in the homonuclear problem with
$\rho_{\rm T}=n_a+2n_m=2$ \cite{Rousseau:Fesh,Rousseau:Mixtures}.  The
spins are modified due to the occupation factors (e.g.\ $S^+=m^\dagger
a a/\sqrt{2}$) but a transition remains.

{\em Conclusions.}--- We have considered the Feshbach problem for two
species of bosons in an optical lattice and have obtained both a rich
phase diagram and the overarching Landau theory.  Within the second
Mott lobe we establish a quantum phase transition described by the
paradigmatic quantum Ising model.  Potential experiments include
magnetization distributions \cite{Lamacraft:OPS}, quantum quenches
\cite{Rossini:Thermal,Calabrese:Quench}, and response near quantum
critical points. Realizing such models on frustrated lattices may
probe connections to dimer models. Finite lifetime effects may also
lead to complex coefficients, and reveal analytic properties such as
the Yang--Lee edge \cite{Fisher:YangLee}.  In the light of these
findings it would be interesting to revisit the numerical simulations
in Refs \cite{Rousseau:Fesh,Rousseau:Mixtures}. In particular, we have
verified that the low {\em excited} states of the bosonic problems are
described by Ising Hamiltonians deep within the Mott
phase. Significantly, the lack of gapless excitations for $g>g_c$, and the
suppression of XY exchange, suggests the absence of novel super-Mott
\cite{Rousseau:Fesh,Rousseau:Mixtures} behavior in this region of the
phase diagram. It would be valuable to explore this.

{\em Acknowledgements.}--- We are extremely grateful to G. Conduit,
F. Essler, J. Keeling, M. K\"ohl, D. Kovrizhin, and S. Powell for
discussions.  MJB, AOS, and BDS acknowledge EPSRC grant
no. EP/E018130/1.  MH was supported by the FWF Schr\"odinger grant
No.~J2583.

\end{document}